\documentclass[aps,prd,showpacs,amssymb,amsmath,amsfonts,superscriptaddress,
twocolumn,floatfix,preprintnumbers,altaffilletter]{revtex4-1}
\usepackage{hyperref}
\hypersetup{
  bookmarksopen=true
}
\usepackage{dcolumn}
\usepackage{caption,subcaption}
\usepackage{graphicx}
\usepackage{threeparttable}
\usepackage{xcolor}
\usepackage{leftidx}
\usepackage[utf8]{inputenc}
\usepackage[export]{adjustbox}
\usepackage[normalem]{ulem}

\newcommand{\iu}{{i\mkern1mu}}

\def\sci#1#2{#1\times10^{#2}}
\def\etal{{\it et al.}}
\def\sign{\textrm{sgn}}

\begin{document}
\title{The Fourier Transform of the Continuous Gravitational Wave Signal}

\author{S.R. Valluri}
\email{valluri@uwo.ca}
\affiliation{Department of Physics and Astronomy, The University of Western Ontario, London, ON N6A 3K7, Canada}
\affiliation{School of Management, Economics and Mathematics, King's University College at Western University, London, Ontario, Canada, N6A 2M3}

\author{V. Dergachev}
\email{vladimir.dergachev@aei.mpg.de}
\affiliation{Max Planck Institute for Gravitational Physics (Albert Einstein Institute), Callinstrasse 38, 30167 Hannover, Germany}
\affiliation{Leibniz Universit\"at Hannover, D-30167 Hannover, Germany}

\author{X. Zhang}
\email{xzha272@uwo.ca}
\affiliation{Department of Statistical and Actuarial Sciences, The
University of Western Ontario, London, ON N6A 3K7, Canada}

\author{F.A. Chishtie}
\email{fachisht@uwo.ca}
\affiliation{Department of Applied Mathematics, The
	University of Western Ontario, London, ON N6A 3K7, Canada}

\begin{abstract}
The direct detection of continuous gravitational waves from pulsars is a much anticipated discovery in the emerging field of multi-messenger gravitational wave (GW) astronomy. Because putative pulsar signals are exceedingly weak large amounts of data need to be integrated to achieve desired sensitivity. Contemporary searches use ingenious ad-hoc methods to reduce computational complexity. In this paper we provide analytical expressions for the Fourier transform of realistic pulsar signals. This provides description of the manifold of pulsar signals in the Fourier domain, used by many search methods. We analyze the shape of the Fourier transform and provide explicit formulas for location and size of peaks resulting from stationary frequencies. We apply our formulas to analysis of recently identified outlier at 1891.76\,Hz.
\end{abstract}

\maketitle
\noindent

\newcounter{saveeqn}
\newcommand{\alpheqn}{\setcounter{saveeqn}{\value{equation}}%
\stepcounter{saveeqn}\setcounter{equation}{0}%
\renewcommand{\theequation}{\mbox{\arabic{saveeqn}-\alph{equation}}}}
\newcommand{\reseteqn}{\setcounter{equation}{\value{saveeqn}}%
\renewcommand{\theequation}{\arabic{equation}}}

\newcounter{savesub}
\newcommand{\alphsub}{\setcounter{savesub}{\value{subsection}}
\stepcounter{savesub}\setcounter{subsection}{0}
\renewcommand{\thesubsection}{\mbox{\arabic{section}\hspace{2pt}\alph{subsection}}}}
\newcommand{\resetsub}{\setcounter{subsection}{\value{savesub}}
\renewcommand{\thesubsection}{\arabic{subsection}}}

\section{Introduction}
Continuous gravitational waves are an eagerly anticipated but elusive phenomena \cite{keith_review}. Despite a series of searches since early 2000 (in particular \cite{O1AllSky2, allsky3, allsky4, EHO1, lvc_O2_allsky, O2_falcon, EHO2}) there have been no loud detections. Some recent papers have seen signals with moderately high SNR, but it is not known yet whether they are due to the instrumental noise or astrophysical signals. 

Continuous gravitational waves are expected from rapidly rotating neutron stars, as well as from more exotic sources \cite{Brito:2017zvb,boson1,boson2,boson3,Horowitz:2019aim,Horowitz:2019pru}.

In this paper we study the Fourier transform of the continuous wave signal using analytical techniques. This study was motivated by loosely coherent algorithms \cite{loosely_coherent1, loosely_coherent2, loosely_coherent3} which adapt to the shape of the signal manifold. 

The Fourier transform of continuous wave signal is analogous to the time-domain representation of a binary waveform. Understanding it is essential for interpreting detection candidates.

The Fourier transform can be computed numerically by first generating gravitational wave signal. Present day gravitational wave detectors produce data at 16384\,Hz sample rate, so a 3-day signal takes $\approx$33\,GB of memory to store. One can reduce storage requirements by heterodyning, but that still results in cumbersome memory and computing requirements.

Our analytical results yield a simple method for determining location and strengths of peaks in the Fourier transform of a continuous wave signal, without the need to generate the entire waveform. This has immediate applications for understanding the influence of detector artifacts. 

Currently available gravitational wave data has frequency spectrum contaminated with numerous sharp peaks \cite{lvc_O2_allsky, O1AllSky2}. The question of overlap of astrophysical signals with these artifacts can be partitioned into direct and inverse problems:

\begin{itemize}
 \item 
In a direct problem, we know the signal parameters with some tolerance and we would like to find out which instrumental lines are located in the signal spectrum. 
\item In the inverse problem, we want to know which signals have spectrum covering a known line.

\end{itemize}

We provide explicit formulas describing location and strength of the peaks in the Fourier transform. These formulas can then be used to efficiently solve both direct and inverse problems of correspondence between signals and sharp detector artifacts. We developed an algorithm for peak computation, detailed in figure \ref{fig:algorithm}.

\section{Signal model}\label{sec:signal_model}

A pure monochromatic signal has linear phase evolution. While this would be computationally simple to search for, the search would be challenging due to confusion of putative signals with numerous instrumental lines \cite{O1AllSky2}.

Realistic gravitational wave signals have multiple sources of modulation, due to Doppler shifts from detector motion relative to the source, possible source motion due to nearby astrophysical bodies, or intrinsic evolution of the source, such as slow decrease in frequency due to energy loss.

All such signals are nearly monochromatic and can be described by the equation
\begin{equation}
h(t)=\Re(a(t) e^{i\phi(t)})
\end{equation}

The Fourier transform of $h(t)$ is thus a convolution of the Fourier transform of amplitude modulation and the phase modulation terms. The amplitude modulation $a(t)$ varies slowly, and its Fourier transform has only five terms: a constant offset, and harmonics (sinusoidal terms) corresponding to periods of half and full sidereal day. Thus most of the complexity is in the behavior of the phase modulation term.

To simplify exposition we assume that $a(t)$ is unity everywhere the signal is defined and focus on the phase modulation terms alone. We do study  the case when input data has gaps, which act as a much stronger amplitude modulation of the waveform.

Such gaps occur naturally due to lock losses in interferometer operations. They can also arise effectively in data that would normally be deweighted due to high noise, or due to unfavorable interferometer angle to the incoming linearly polarized signal.

The phase modulation $\phi(t)$ is a powerful tool in separating astrophysical signals from detector artifacts, but its complicated form and dependence on many parameters, such as source and detector locations and frequency drift parameters presents a computational challenge.

It is instructive to consider a simplified situation of a fixed frequency source and a detector following two superimposed circular motions - one around Earth's axis and one of Earth around the Sun.

First,  we compute relative detector position to Earth center:
\begin{equation}
\vec{r}_{\textrm{Earth}}= \left(\begin{array}{l}
                                \cos(\omega_{\textrm{rot}}t+\alpha_0)\cos(\delta_0) \\
                                 \sin(\omega_{\textrm{rot}}t+\alpha_0)\cos(\delta_0) \\
                                 \sin(\delta_0) \\
                                \end{array}\right)
R_{\textrm{Earth}}
\end{equation}
where $\alpha_0$ and $\delta_0$ are the detector longitude and latitude locations correspondingly.

The full motion of the detector is then described as
\begin{equation}
\vec{r}_{\textrm{det}}=\vec{r}_{\textrm{Earth}}+\vec{v}\cos(\omega_{\textrm{orb}}t)+\vec{u}\sin(\omega_{\textrm{orb}}t)
\end{equation}
where $\vec{u}$ and $\vec{v}$ are two perpendicular vectors in the ecliptic plane parametrizing Earth's orbital motion.

This can be generalized as
\begin{equation}
\label{eqn:rdet}
\vec{r}_{\textrm{det}}=\vec{r}_{\textrm{off}}+\sum_{k=1}^K \vec{v}_k\cos(\omega_kt)+\vec{u}_k\sin(\omega_{k}t)
\end{equation}
Here $\vec{r}_{\textrm{off}}$ is a constant offset, which for circular approximation is $R_{\textrm{Earth}}\sin(\delta_0) \hat{\vec{z}}$. Since the offset is constant, it only affects absolute signal phase. For simplicity we will assume $\vec{r}_{\textrm{off}}=0$ in subsequent calculations.

Our simplified example with two modulations arising from circular motions corresponds to $K=2$. The equation \ref{eqn:rdet} is general enough that one can fit any realistic signal by including additional harmonics, for example, due to planetary perturbations. The algorithm presented in Figure \ref{fig:algorithm} is run using solar system barycenter timings incorporating full complexity of the underlying signal.

The direction to the source is given by
\begin{equation}
\hat{n}_{\textrm{source}}=\left(\begin{array}{l}
                                 \cos(\alpha)\cos(\delta) \\
                                 \sin(\alpha)\cos(\delta) \\
                                 \sin(\delta) \\
                                \end{array}\right)
\label{eq:nsource}
\end{equation}
Here $\alpha$ is the right ascension in radians [$0$, $2\pi$) and $\delta$ is the declination in radians [$-\pi/2$, $\pi/2$].

The detector velocity vector is 
\begin{equation}
\vec{v}_{\textrm{det}}=\frac{d}{dt}\vec{r}_{\textrm{det}}=\sum_{k=1}^K \vec{u}_k \omega_k \cos(\omega_kt)-\vec{v}_k\omega_{k}\sin(\omega_{k}t)
\end{equation}

The Doppler shift is computed from the formula:
\begin{equation}
\begin{split}
\mathcal{D}&=\frac{\hat{n}_{\textrm{source}}\cdot\vec{v}_{\textrm{det}}}{c}\\
&=\sum_k \frac{\hat{n}_{\textrm{source}}\cdot\vec{u}_k }{c} \omega_k \cos(\omega_kt)-\frac{\hat{n}_{\textrm{source}}\cdot\vec{v}_k}{c}\omega_{k}\sin(\omega_{k}t)
\end{split}
\end{equation}

We introduce relative modulation depth $a_k^\textrm{rel}$ and modulation phase $\phi_k$:
\begin{equation}
a_k^\textrm{rel}=\frac{\omega_k}{c}\sqrt{ (\hat{n}_{\textrm{source}}\cdot\vec{u}_k)^2 + (\hat{n}_{\textrm{source}}\cdot\vec{v}_k)^2} 
\end{equation}

\begin{equation}
\phi_k=\arctan\left(\frac{\hat{n}_{\textrm{source}}\cdot\vec{u}_k}{\hat{n}_{\textrm{source}}\cdot\vec{v}_k}\right)
\end{equation}

Then the Doppler shift becomes:
\begin{equation}
\begin{split}
\mathcal{D}&=\sum_{k=1}^K a_k^\textrm{rel} \cos(\omega_kt+\phi_k)
\end{split} 
\end{equation}

Let us widen the model for our source signal to include polynomial frequency evolution, which is observed in radio pulsars:
\begin{equation}
f(t)=\sum_{n=0}^{N}f_n \frac{t^n }{n!}
\end{equation}
The signal received at the detector is then
\begin{equation}
f(t)=\left(\sum_{n=0}^{N}f_n \frac{t^n }{n!}\right)\left(1+\sum_{k=1}^K a_k^\textrm{rel} \cos(\omega_kt+\phi_k)\right)
\end{equation}
This ignores relativistic corrections.

Let us assume that the products $f_na_k^{\textrm{rel}}$ are negligible for all $n\geq1$. This is the case for frequency drifts in most known radio pulsars. For example, with $f_1\approx 10^{-12}$\,Hz/s and with Doppler shifts from Earth orbital motion $a^{\textrm rel}_2 \approx 10^{-4}$ we find the product $f_1 a^{\textrm rel}_2 \approx 10^{-16}$\,Hz/s much smaller than resolution of most all-sky searches.
Then the signal model simplifies to:
\begin{equation}
f(t)=\sum_{n=0}^{N}f_n \frac{t^n }{n!}+f_0\sum_{k=1}^K a_k^\textrm{rel} \cos(\omega_kt+\phi_k)
\end{equation}

We now introduce phase modulation depth $a_k$ as
\begin{equation}
a_k=\frac{2\pi f_0a_k^\textrm{rel}}{\omega_k}=\frac{2\pi f_0}{c}\sqrt{ (\hat{n}_{\textrm{source}}\cdot\vec{u}_k)^2 + (\hat{n}_{\textrm{source}}\cdot\vec{v}_k)^2} 
\end{equation}

Then the phase model of our signal is
\begin{equation}
\phi(t)=\phi_0+2\pi\sum_{n=1}^{N}f_{n-1} \frac{t^n }{n!}+\sum_{k=1}^K a_k \sin(\omega_kt+\phi_k)
\end{equation}
Here $\phi_0$ controls the initial phase of the signal.

To make sense of modulation amplitudes and phases, we focus on our initial case of two circular modulations.

The vectors $u_1$ and $v_1$ describing Earth's rotation are
\begin{equation}
\vec{v}_1=R_{\textrm{Earth}}\left(\begin{array}{l}
                                  \cos \alpha_0 \cos \delta_0 \\
                                  \sin \alpha_0 \cos \delta_0 \\
                                  0
                                  \end{array}
\right)
\end{equation}
\begin{equation}
\vec{u}_1=R_{\textrm{Earth}}\left(\begin{array}{l}
                                  -\sin \alpha_0 \cos \delta_0 \\
                                  \cos \alpha_0 \cos \delta_0 \\
                                  0
                                  \end{array}
\right)
\end{equation}
where $\alpha_0$ and $\delta_0$ are detector longitude and latitude correspondingly.

Then the parameters corresponding to Earth's rotation are
\begin{equation}
\begin{array}{l}
\phi_1=\arctan\left(\frac{\hat{n}_{\textrm{source}}\cdot\vec{u}_k}{\hat{n}_{\textrm{source}}\cdot\vec{v}_k}\right)=\\
\quad\quad=\arctan\left(\frac{\sin(\alpha-\alpha_0)\cos(\delta_0)}{\cos(\alpha-\alpha_0)\cos(\delta_0)}\right)=\alpha-\alpha_0
\end{array}
\end{equation}
\begin{equation}
a_1= \frac{2\pi f_0R_{\textrm{Earth}}}{c}\left|\cos(\delta)\cos(\delta_0)\right|
\end{equation}

We see that modulation phase $\phi_1$ is just the difference between source right ascension and detector longitude.

The vectors $u_2$ and $v_2$ describing Earth orbital motion are
\begin{equation}
\vec{v}_2=R_{\textrm{orb}}\left(\begin{array}{l}
                                  1 \\
                                  0 \\
                                  0
                                  \end{array}
\right)
\end{equation}
\begin{equation}
\vec{u}_2=R_{\textrm{orb}}\left(\begin{array}{l}
                                  0 \\
                                  \cos\epsilon \\
                                  \sin\epsilon
                                  \end{array}
\right)
\end{equation}
where $\epsilon=23.4^\circ$ is the obliquity of the ecliptic

The parameters corresponding to Earth orbital motion around the Sun are somewhat more complicated:
\begin{equation}
\phi_2=\arctan\left(\frac{\sin(\alpha)\cos(\delta)\cos(\epsilon)+\sin(\delta)\sin(\epsilon)}{\cos(\alpha)\cos(\delta)}\right)
\end{equation}
\begin{equation}
\begin{array}{l}
\displaystyle
a_2=\frac{2\pi f_0R_{\textrm{orb}}}{c} \cdot\\
\cdot \sqrt{\cos^2(\alpha)\cos^2(\delta)+(\sin(\alpha)\cos(\delta)\cos(\epsilon)+\sin(\delta)\sin(\epsilon))^2}
\end{array}
\end{equation} 

This complexity is due to the choice of equatorial coordinate system. Had we chosen  ecliptic coordinates instead the orbital motion parameters would be simple, while the Earth rotation parameters have similar expressions to the above, as we will essentially exchange indices. As we will see later the shorter period motion introduces more complexity in the Fourier transform, so it makes sense to use the equatorial coordinate system in applications.

\section{Fourier transform of quasi-monochromatic signal}

\subsection{Signal spectrum}

In the general case the signal spectrum is 

\begin{equation}
\begin{array}{l}
\displaystyle
\tilde{h}_0(\mathit{f}) =\int_{-T/2}^{T/2}\exp\left(\iu \phi(t) \right)e^{-\iu2\pi\mathit{f}t}dt =\\
\displaystyle
\quad\quad=\int_{-T/2}^{T/2}\exp\left\{\iu \phi_0+2\pi \iu\sum_{n=2}^{N}f_{n-1} \frac{t^n }{n!}+\right.\\
\displaystyle
\quad\quad\quad\quad+\left.\iu\sum_{k=1}^K a_k \sin(\omega_kt+\phi_k) \right\} \cdot e^{-\iu2\pi\left(\mathit{f}-f_0\right)t}dt\\
\end{array}
\end{equation}

Thus the spectrum depends on initial signal phase $\phi_0$, initial frequency $f_0$, higher order frequency expansion parameters $f_k$ (for $k\ge 1$), phase modulation depth $a_k$ and modulation phase $\phi_k$. 

For searches less than 30 days the effect of third order and higher frequency derivatives can be neglected for astrophysical sources. Keeping terms up to a second order in frequency, the equation simplifies to

\begin{equation}
\begin{array}{l}
\displaystyle
\tilde{h}_0(\mathit{f}) =\int_{-T/2}^{T/2}\exp\left\{\iu \phi_0+\iu 2\pi\left(f_{1} \frac{t^2 }{2} + f_2 \frac{t^3}{6}\right)+ \right.\\
\quad\quad\quad\quad \left. + \iu\sum_{k=1}^K a_k \sin(\omega_kt+\phi_k) \right\}\cdot e^{-\iu2\pi\left(\mathit{f}-f_0\right)t}dt\\
\end{array}
\end{equation}

The treatment of sinusoidal phase modulation can use either the Jacobi-Anger expansion in terms of Bessel functions or approximation of the sine function by polynomials.

\subsection{Polynomial approximation}

The polynomial approximation is particularly effective when $\omega_l T$ is small.

For example:
\begin{equation}
\label{eqn:expansion}
\begin{array}{l}
\displaystyle
a_k\sin(\omega_k t+\phi_k)=a_k\left(\omega_kt-\frac{\omega_k^3t^3}{6}+O(\omega_k^5t^5)\right)\cos(\phi_k)+\\
\quad\quad+a_k\left(1-\frac{\omega_k^2t^2}{2}+O(\omega_k^4t^4)\right)\sin(\phi_k))
\end{array}
\end{equation}

\begin{table}[htbp]
\begin{center}
\begin{tabular}{lrrrr}\hline
Modulation & Source & Earth rotation & Orbital motion & Unit \\
 term  & frequency & & & \\
\hline
\hline
$\omega_l$ & - & $6.3$ &  $0.017$ &  (1/day) \\

$a_l$  & 200\,Hz  & 23  & 630000 &  - \\
$a_l$ & 1000\,Hz  & 115 &  3200000 & - \\
 $a_l$ & 2000\,Hz  & 230 & 6300000  & - \\

$a_l\omega_l$  & 200\,Hz  & 0.0017 & 0.123 & Hz \\
$a_l\omega_l$ & 1000\,Hz  & 0.0084  & 0.63   & Hz\\
 $a_l\omega_l$ & 2000\,Hz  &  0.017 &  1.23 & Hz \\
 
$a_l\omega_l^2$  & 200\,Hz  & $\sci{1.2}{-7}$ & $\sci{2.4}{-8}$ & Hz$^2$ \\
$a_l\omega_l^2$ & 1000\,Hz  &$\sci{6.1}{-7}$ &  $\sci{1.2}{-7}$ & Hz$^2$\\
 $a_l\omega_l^2$ & 2000\,Hz  & $\sci{1.2}{-6}$ &  $\sci{2.4}{-7}$ & Hz$^2$ \\
\hline
\end{tabular}
\end{center}
\caption{Modulation parameters}{Modulation parameters for various sources. The amplitude modulation values are worst case, as seen in LIGO Livingston interferometer. Phase and frequency modulation are dominated by orbital motion, while the frequency derivatives are larger for terms from Earth rotation.}
\label{tab:modulation_parameters}
\end{table} 

Table \ref{tab:modulation_parameters} shows modulation parameters for sources emitting at various example frequencies. 

For example, in the case of $T=3{\textrm{\,days}}$ we find that the Earth's orbital motion is a good candidate for polynomial expansion and would need terms up to a cubic order.

Indeed, the error in equation \ref{eqn:expansion} can be bounded by the 4-th order term:
\begin{equation}
a_l \frac{\omega_l^4T^4}{4!2^4} \le 0.11
\end{equation}
Here we assumed the expansion is centered on the middle of the interval so the maximum time is $T/2$.

Let $L_E$ be the set of indices describing expanded harmonics.
Consider the following integral by neglecting constant phase term:
\begin{equation}
\label{eqn:approximation1}
 \begin{array}{l}
 \displaystyle
\tilde{h}_0(\mathit{f}) =\int_{-T/2}^{T/2}\exp\left(2\pi \iu g_{1} \frac{t^2 }{2}+2\pi \iu g_{2} \frac{t^3 }{6}+\right.\\
\displaystyle
 \quad\quad \left.+\iu\sum_{k\notin L_E} a_k \sin(\omega_kt+\phi_k) \right)\cdot e^{-\iu2\pi\left(\mathit{f}-g_0\right)t}dt\\
\end{array}
\end{equation}
where coefficients $g_n$ have been introduced that describe both initial polynomial frequency modulation parameters and the contribution from polynomial expansion of 
sinusoidal modulations:
\begin{equation}
\displaystyle
g_n=f_n+\sum_{k\in L_E}\frac{a_k \omega_k^{n+1}}{2\pi (n+1)!}\cos\left(\phi_k+\frac{\pi n}{2}\right)
\end{equation}
This model of polynomial plus harmonics is very effective in describing a realistic pulsar signal. As we will show later, the phase behavior of exact Solar System barycenter timings can be approximated with a single harmonic plus a third order polynomial over any data stretch of $3$\,days or less. 

\subsection{Jacobi-Anger expansion}

We now focus on the application of the Jacobi-Anger expansion. Applying it to all sinusoidal terms we get:
\begin{equation}
 \begin{array}{l}
 \displaystyle
\tilde{h}_0(\mathit{f}) =\int_{-T/2}^{T/2}\exp\left( 2\pi \iu g_{1} \frac{t^2 }{2}+2\pi \iu g_{2} \frac{t^3 }{6}-\iu2\pi\left(\mathit{f}-g_0\right)t\right)\cdot\\
  \displaystyle
\quad\quad\quad\cdot\exp\left(\iu\sum_{l\notin L_E} a_l \sin(\omega_lt+\phi_l) \right)dt=\\
 \displaystyle
\quad\quad=\int_{-T/2}^{T/2}\exp\left( 2\pi \iu g_{1} \frac{t^2 }{2}+2\pi \iu g_{2} \frac{t^3 }{6}-\iu2\pi\left(\mathit{f}-g_0\right)t\right)\cdot\\
 \displaystyle
\quad\quad\quad\cdot \prod_{l\notin L_E} \sum_{k_l} \iu^{k_l}J_{k_l}(a_l)\exp(\iu k_l(\omega_lt+\phi_l))dt
\end{array} 
\end{equation}

The product and sum symbols can be exchanged yielding a sum over multi-indices $\vec{k}=(k_1,\dots,k_M)$:
\begin{equation}
 \begin{array}{l}
 \displaystyle
\tilde{h}_0(\mathit{f}) =\int_{-T/2}^{T/2}\exp\left( 2\pi \iu g_{1} \frac{t^2 }{2}+2\pi \iu g_{2} \frac{t^3 }{6}-\iu2\pi\left(\mathit{f}-g_0\right)t\right)\cdot\\
 \displaystyle
\quad\quad\quad\cdot \sum_{\vec{k}} \prod_{l\notin L_E}  \iu^{k_l}J_{k_l}(a_l)\exp(\iu k_l(\omega_lt+\phi_l))dt
\end{array} 
\end{equation}
The indices $\vec{k}$ span an infinite lattice for exact expression. However, the values $J_{k_l}(a_l)$ decrease rapidly for $k_l\gg |a_l|$, allowing a finite sum to be used in practical calculations.

The number of remaining indices in a sum depends on modulation depth $a_l$ and can be fairly substantial even for relatively small modulation values. This complexity is intrinsic to the problem, as can be confirmed by examining numerically computed Fourier transform in figure \ref{fig:spectrum_support1} - the multitude of peaks would need separate harmonic terms to produce them.

Mathematically, this can be understood as follows. 

First we perform the expansion of longer period harmonics as done in Equation \ref{eqn:approximation1} keeping only one remaining harmonic. Then we split the integral into pieces of length matching one period $T_p$ (where $T_p\omega_1=2\pi$). We assume the full integration interval is the integer multiple of period $T_p$:

\begin{equation}
 \begin{array}{l}
 \displaystyle
\tilde{h}_0(\mathit{f}) =\sum_{m=0}^M \int_{-T_p/2+mT_p}^{T_p/2+mT_p} \cdot\\ 
\displaystyle
\quad\quad \cdot \exp\left(2\pi \iu g_{1} \frac{t^2 }{2}+2\pi \iu g_{2} \frac{t^3 }{6}+\iu a_1 \sin(\omega_1t+\phi_1) \right)\cdot\\
 \quad\quad\quad\quad \cdot e^{-2\pi \iu\left(\mathit{f}-g_0\right)t}dt\\
\end{array}
\end{equation}
Shifting the internal integration variable by $mT_p$ we obtain:
\begin{equation}
 \begin{array}{l}
 \displaystyle
\tilde{h}_0(\mathit{f}) =\sum_{m=0}^M \int_{-T_p/2}^{T_p/2} \exp\left(2\pi \iu g_{1} \frac{(t+mT_p)^2 }{2}+\right.\\
\quad\quad +2\pi \iu g_{2} \frac{(t+mT_p)^3 }{6} 
\displaystyle
\left. \vphantom{\frac{T}{2}}+ \iu  a_1 \sin(\omega_1t+\phi_1) \right)\cdot\\
\quad\quad \cdot e^{-2\pi \iu\left(\mathit{f}-g_0\right)(t+mT_p)}dt\\
\end{array}
\end{equation}
The argument of the sine function is unmodified because we shift by integral number of periods.

The Taylor formula provides a convenient way to compute a shift of any analytic function:
\begin{equation}
p(t+T)=\sum_{n=0}^{\infty} \frac{T^n}{n!} \frac{d^n}{dt^n} p(t) 
\end{equation}
For polynomials the sum is finite because higher order derivatives vanish.
Our polynomial is only third order:
\begin{equation}
p(t)=2\pi (g_0-f)t+2\pi g_1 \frac{t^2}{2}+2\pi g_2 \frac{t^3}{6}
\end{equation}
Leading to a simple expression for the shift:
\begin{equation}
\begin{array}{ll}
p(t+mT_p)= & p(t)+   \\ 
 & + mT_p 2\pi(g_0-f +g_1 t+g_2 \frac{t^2}{2})+ \\
 & + \frac{m^2T_p^2}{2} 2\pi (g_1+g_2 t)+\\
 & + \frac{m^3T_p^3}{6} 2\pi g_2
\end{array}
\end{equation}
Thus the shifted integral can be described as a convolution of a single-period Fourier transform with a Fourier transform of $\exp(\iu (p(t+mT_p)-p(t))$. The latter can be separated into three parts:
\begin{itemize}
 \item a multiplication by the phase 
 \begin{equation}
\exp\left(2\pi \iu \left(\frac{g_1m^2T_p^2}{2}+\frac{g_2m^3T_p^3}{6}\right)\right)  
 \end{equation}
 \item a shift in frequency by $g_1mT+g_2 m^2 T_p^2/2$ which we denote by operator $\mathbb S(m)$.
 \item and a convolution with Fourier transform of a Gaussian $\exp(\pi \iu T_pg_2 t^2)$ iterated $m$ times. We denote a single iteration of the convolution by operator $\mathbb T$.
\end{itemize}

Then the full integral can be expressed as

\begin{equation}
\label{eqn:time_lattice}
 \begin{array}{l}
 \displaystyle
\tilde{h}_0(\mathit{f}) =\sum_{m=0}^M e^{2\pi \iu \left(\frac{g_1m^2T_p^2}{2}+\frac{g_2m^3T_p^3}{6}\right)} {\mathbb T}^m {\mathbb S}(m) \,\tilde{h}_0^1(\mathit{f}) \\
\end{array}
\end{equation}
where $\tilde{h}_0^1(\mathit{f})$ denotes a single period Fourier transform:
\begin{equation}
\begin{array}{l}
\displaystyle
 \tilde{h}_0^1(\mathit{f})=\int_{-T_p/2}^{T_p/2}
 \exp\left(2\pi \iu g_{1} \frac{t^2 }{2}+2\pi \iu g_{2} \frac{t^3 }{6}+\right.\\
 \displaystyle 
\quad\quad\left.\vphantom{\frac{T}{2}}+\iu a_1 \sin(\omega_1t+\phi_1) \right)
 e^{-\iu2\pi\left(\mathit{f}-g_0\right)t}dt\\
 \end{array}
\end{equation}

This expression explains features of the Figure \ref{fig:spectrum_support1}. The repeated pattern is due to iterations of the operator $\mathbb{T}$ and frequency shift $\mathbb S(m)$. However, both this operator and the frequency shift $\mathbb{S}$ introduce frequency shifts that are not aligned to frequency bins of the full Fourier transform. Thus the height of peaks varies with each iteration. 

The variation in phase together with convolution acts to scramble the heights of smaller peaks resulting in a signature of the underlying signal.

\section{Fourier transform shape}
The equation \ref{eqn:time_lattice} allows us to understand the Fourier transform of continuous wave signals in a qualitative way. For practical applications it is desirable to know the details such as location of the peaks and their heights. 

While this can be done by the numerical integration of formula \ref{eqn:time_lattice}, the computation is comparable in difficulty to taking the Fourier transform directly. What we would like instead is a simple formula depending on parameters of the signal $g_k$, $\omega_1$ and $\phi_1$.

To obtain such formulas, consider the integral

\begin{equation}
 \begin{array}{l}
 \displaystyle
\tilde{h}_0(\mathit{f})=\int_{-T/2}^{T/2}\exp\left( 2\pi \iu g_{1} \frac{t^2 }{2}+2\pi \iu g_{2} \frac{t^3 }{6}+  \right.\\
\displaystyle 
\quad\quad \left.\vphantom{\frac{T}{2}}+\iu\sum_{l\notin L_E} a_l \sin(\omega_lt+\phi_l)  \right)e^{-2\pi \iu\left(\mathit{f}-g_0\right)t} dt=\\
\displaystyle 
\quad\quad =\int_{-T/2}^{T/2}\exp\left(\iu \Lambda(t)\right) dt
\end{array}
\end{equation}
Because of the imaginary terms in the exponent it is highly oscillatory. These oscillations will cancel out (on average), except in points where derivative of $\Lambda(t)$ vanishes:

\begin{equation}
 \displaystyle
\Lambda'(t)=  2\pi g_{1}t+2\pi  g_{2} \frac{t^2 }{2}+\sum_{l\notin L_E} a_l\omega_l \cos(\omega_lt+\phi_l)-2\pi\left(\mathit{f}-g_0\right)
\end{equation}

This can be rewritten as
\begin{equation}
\label{eqn:spectrum_support}
 \displaystyle
F(t)=g_0+g_{1}t+g_{2} \frac{t^2 }{2}+\sum_{l\notin L_E} \frac{a_l\omega_l}{2\pi} \cos(\omega_lt+\phi_l)=f
\end{equation}

\begin{figure}[htbp]
\begin{center}
   \includegraphics[width=3.6in]{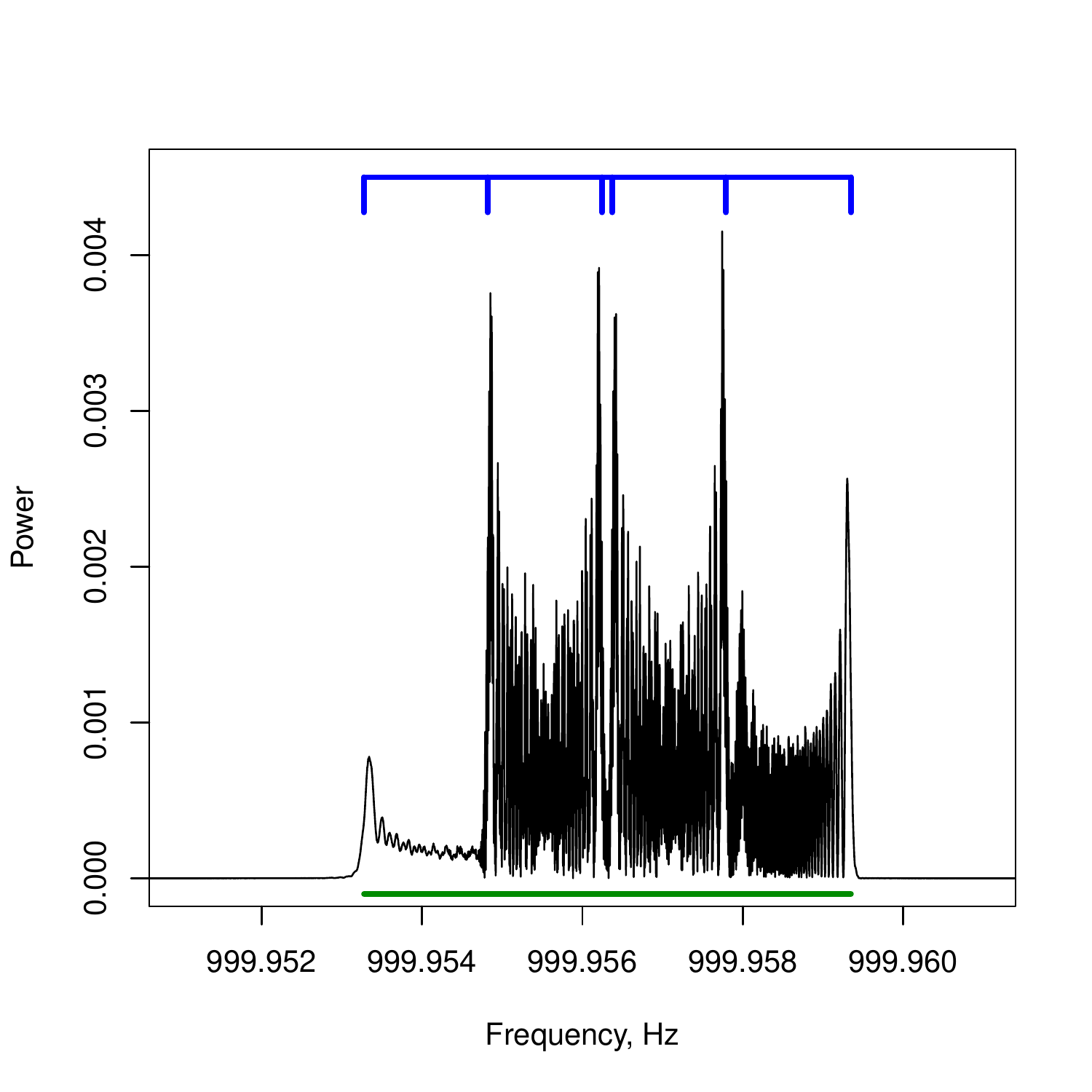}
 \caption{Example 3-day Fourier transform of 1000\,Hz monochromatic signal from source at right ascension $0^\circ$ and declination $0^\circ$. The green line at the bottom of the plot shows spectrum support region estimated using equation \ref{eqn:spectrum_support}. The blue line at the top shows peak locations estimated using equation \ref{eqn:peak_locations}. The Fourier transform was computed assuming 100\% duty cycle for LIGO Hanford interferometer. The 3-day segment started at GPS 1160657033.}
\label{fig:spectrum_support1}
\end{center}
\end{figure}

Because $f$ is a free parameter, the support of the spectrum of our signal is close to the image of the interval $[-T/2, T/2]$ under a function $F(t)$ (Figure \ref{fig:spectrum_support1}).

The largest peaks in the spectrum should correspond to the values of $f$ for which larger time intervals have stationary phase, and thus to the points $f_a=F(t_a)$ such that  the derivative of $F$ vanishes:
\begin{equation}
 \displaystyle
\label{eqn:spectrum_support_deriv}
F'(t_a)=g_{1}+g_{2}t_a-\sum_{l\notin L_E}\frac{a_l\omega_l^2}{2\pi} \sin(\omega_lt_a+\phi_l)=0
\end{equation}

This equation has an approximate solution in the special case of a single sinusoidal term and small parameters $g_{1}$ and $g_{2}$. 

In this case the equation reduces to
\begin{equation}
 \displaystyle
\sin(\omega_1t_a+\phi_1)=\frac{2\pi g_{1}}{a_l\omega_1^2}+\frac{2\pi g_{2}}{a_l\omega_1^2}t_a 
\end{equation}

Let $t_a^0=\frac{-\phi_1+\pi n}{\omega_1}$ be the zero of the sine function. 

Applying one step of Newton-Raphson method to find the solution of the above equation using $t^0_a$ as the initial value, we have

\begin{equation}
\label{eqn:stationary_points}
    t_a = t^0_a - \frac{g_{1}+g_{2}t^0_a}{g_{2}-(-1)^n\frac{a_1\omega_1^3}{2\pi}}
\end{equation}

Let us check how close we got to true zero of $F'(t)$. We substitute $t_a$ into equation \ref{eqn:spectrum_support_deriv}:

\begin{equation}
 \begin{array}{l}
 \displaystyle
F'(t_a)=g_{1}+g_{2}t_a-\frac{a_l\omega_1^2}{2\pi} \sin(\omega_1 t_a+\phi_l)\approx\\
\quad\quad\quad \approx a_1\omega_1^2 O\left(\left(-\omega_1 \frac{g_{1}+g_{2}t^0_a}{g_{2}-(-1)^n\frac{a_1\omega_1^3}{2\pi}} \right)^3\right)
\end{array}
\end{equation}

We see the approximate solution $t_a$ has canceled all linear terms.

To find out the frequencies of the peaks we  can now substitute $t_a$ into the Eq. \ref{eqn:spectrum_support}.  We find

\begin{equation}
\label{eqn:peak_locations}
\begin{array}{l}
\displaystyle
f \approx F(t_a) \approx g_0+g_{1}t_a+g_{2} \frac{t_a^2}{2}+ \\
\quad\quad\quad+(-1)^n\frac{a_1\omega_1}{2\pi} \left(1 - \frac{1}{2}\left(\frac{\omega_1(g_{1}+g_{2}t^0_a)}{g_{2}-(-1)^n\frac{a_1\omega_1^3}{2\pi}}\right)^2\right)
\end{array}
\end{equation}

The zeros of the second frequency derivative are simpler to find:
\begin{equation}
 \displaystyle
\label{eqn:spectrum_support_deriv2}
F''(t_a)=g_{2}-\sum_{l\notin L_E}\frac{a_l\omega_l^3}{2\pi} \cos(\omega_lt_b+\phi_l)=0
\end{equation}
In the case of a single sinusoidal term we have:
\begin{equation}
\frac{a_1\omega_1^3}{2\pi}\cos(\omega_1 t_b +\phi_1)=g_2
\end{equation}

\begin{equation}
\label{eqn:inflection_locations1}
t_b=\frac{\arccos\left(\frac{2\pi g_2}{a_1\omega_1^3}\right)-\phi_1+\pi n}{\omega_1} 
\end{equation}

For the common case of $\left|2\pi g_2\right|\ll \left|a_1\omega_1^3\right|$ the formula simplifies to 

\begin{equation}
\label{eqn:inflection_locations2}
t_b=-\phi_1+\pi n 
\end{equation}

\begin{figure}[htbp]
\begin{center}
   \includegraphics[width=3.6in]{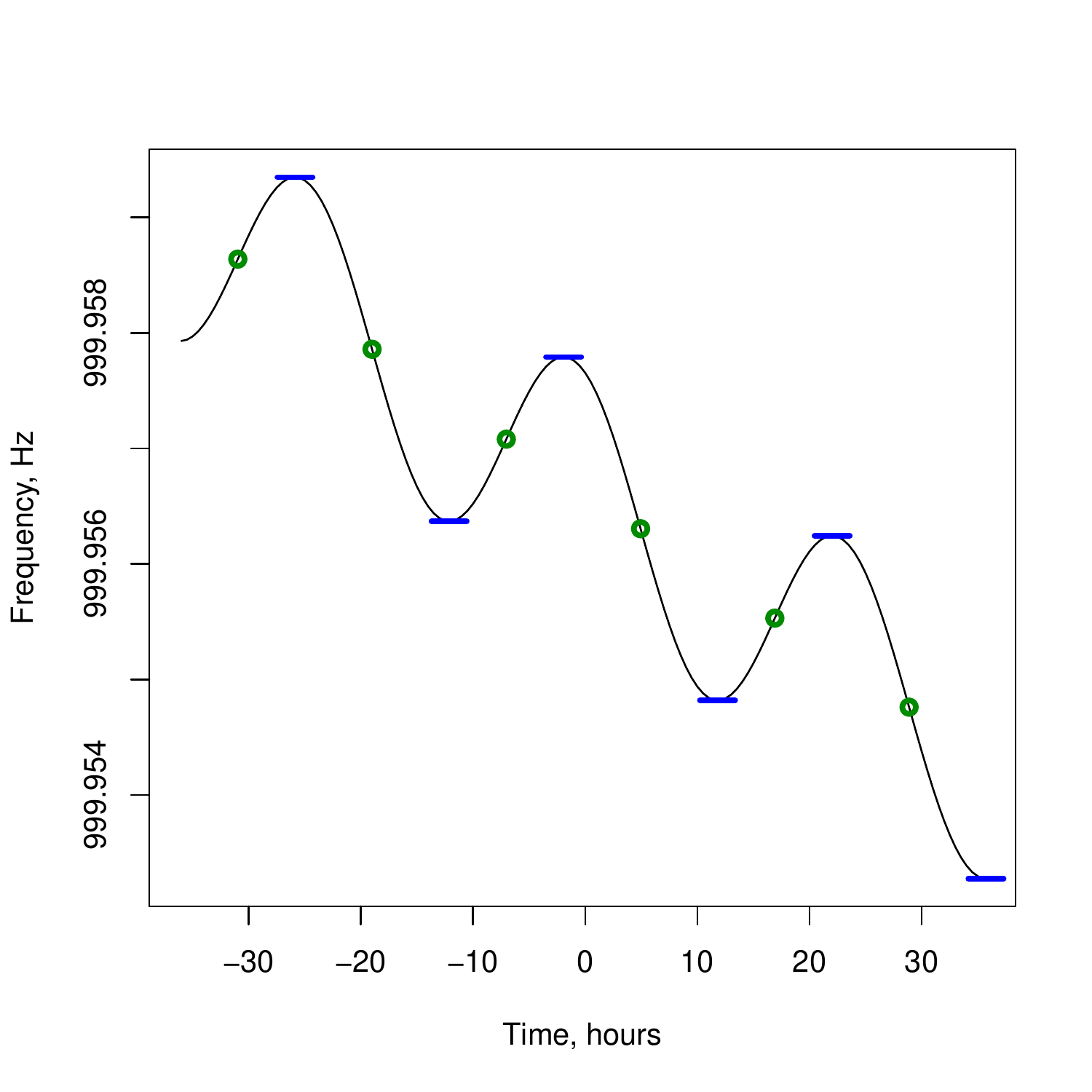}
 \caption{Example 3-day frequency evolution of 1000\,Hz monochromatic signal from source at right ascension $0^\circ$ and declination $0^\circ$. The blue lines mark locations of local frequency minima and maxima. Green circles mark location of inflection points. The frequencies were computed for LIGO Hanford interferometer. The 3-day segment started at GPS 1160657033.}
\label{fig:frequency_evolution}
\end{center}
\end{figure}

To test these formulas we generated barycentered time series for LIGO Hanford and Livingston interferometers \cite{aLIGO} for multiple sky locations over one year period. 

Figure \ref{fig:frequency_evolution} shows locations of local frequency maxima and minima, as well as inflection points where the second frequency derivative vanishes for a portion of this data for a 3-day period starting at GPS time 1160657033 generated for LIGO Hanford interferometer.

Figure \ref{fig:spectrum_support1} shows power spectrum (absolute value squared of the Fourier transform) of a pure $1000$\,Hz signal with amplitude $1$ as observed by LIGO Hanford interferometer. Unlike real data sets which have gaps due to interferometer lock loss this time series is contiguous.

We observe that the spectrum support (marked by the thick green line below the graph) is correctly computed by formula \ref{eqn:spectrum_support}. 

The peak locations marked by short blue lines at the top of the graph were computed with formula \ref{eqn:peak_locations} and correspond well with numerical results.

Having found peak locations we would like to have a measure of their heights, as those clearly vary. 

Near a point of stationary frequency the Fourier transform has the form 
\begin{equation}
h_{\textrm{local}}(f)=\int_{t_0}^{t_1} e^{2\pi i \left(\phi+F(t_a) (t-t_a)+\tilde{g}_2 (t-t_a)^3/6 \right)}  e^{-2\pi i f t}dt 
\end{equation}
where we introduced $\tilde{g}_2$:
\begin{equation}
\tilde{g}_2=g_2-\frac{a_1}{2\pi}\frac{\cos\left(\omega_1 t_a+\phi_1\right)}{6} 
\end{equation}

This equation is designed to describe the vicinity of $f=F(t_a)$. The limits of the integration $t_0$ and $t_1$ bound the region where the approximation holds, in particular there is no need to integrate over points close to other stationary frequency points.

The height of the peak is given by
\begin{equation}
\label{eqn:peak_height1}
|h_{\textrm{local}}(F(t_a))|=\left|\int_{t_0}^{t_1} e^{2\pi i \tilde{g}_2 (t-t_a)^3/6}dt \right| 
\end{equation}
We now need to find out which values of $t_0$ and $t_1$ to use. Naively we might expect that one should use a small interval where the frequency does not change far away from stationary value $F(t_a)$. 

However, this will grossly underestimate peak height. The reason is that the value of truncated Airy function (equation \ref{eqn:peak_height1}) keeps growing with increasing time interval, as nearby frequencies contribute due to spectral leakage.

A good heuristic is to choose $t_0$ and $t_1$ to be the inflection points, or a data boundary if it occurs earlier. 

The truncated Airy function has an expression in terms of incomplete Gamma function:
\begin{equation}
\label{eqn:Gamma_diff1}
\int_{t_0}^{t_1}e^{it^3} dt=\frac{1}{3\sqrt[3]{-\iu }}\left(\Gamma\left(\frac{1}{3}, -\iu t_0^3\right)-\Gamma\left(\frac{1}{3}, -\iu t_1^3\right)\right)
\end{equation}

This equation has some ambiguity as to the branch of cubic roots. This arises purely from using the incomplete $\Gamma$ function:
\begin{equation}
\Gamma(a, z)=\Gamma(a)\left(1-z^a e^{-z}\sum_{k=0}^{\infty}\frac{z^k}{\Gamma(a+k+1)}\right) 
\end{equation}
The constant terms in the formula above subtract when substituted in Eq. \ref{eqn:Gamma_diff1}.

This can also be seen by expanding $e^{it^3}$ into a Taylor series and integrating the result:
\begin{equation}
\int_{t_0}^{t_1}e^{it^3} dt=\sum_{k=0}^{\infty} \frac{i^k t_1^{3k+1}}{k!(3k+1)}-\sum_{k=0}^{\infty} \frac{i^k t_0^{3k+1}}{k!(3k+1)}
\end{equation}

However, for practical application it is convenient to approximate with a heuristic piece-wise linear function that captures the general shape of the integral.

To do this, we introduce the function 
\begin{equation}
H(a)=\left\{
\begin{array}{ll}
|a|  &\quad \textrm{ when }|a|<0.4   \\
0.4  &\quad \textrm{ when }|a|\ge 0.4
\end{array}
\right. 
\end{equation}

Then 
\begin{equation}
\label{eqn:peak_height2}
\begin{array}{l}
\displaystyle
|h_{\textrm{local}}(F(t_a))|=\left|\int_{t_0}^{t_1} e^{2\pi i \tilde{g}_2 (t-t_a)^3/6}dt \right| \approx \\
\displaystyle
\quad\quad\quad\approx \frac{1}{\kappa} \left|H(t_1\kappa)-(-1)^{\sign(t_0)\sign(t_1)}H(t_0\kappa)\right|
\end{array}
\end{equation}
where $\kappa=\sqrt[3]{6/\tilde{g}_2}$.

\begin{figure}[htbp]
\begin{center}
   \includegraphics[width=3.6in]{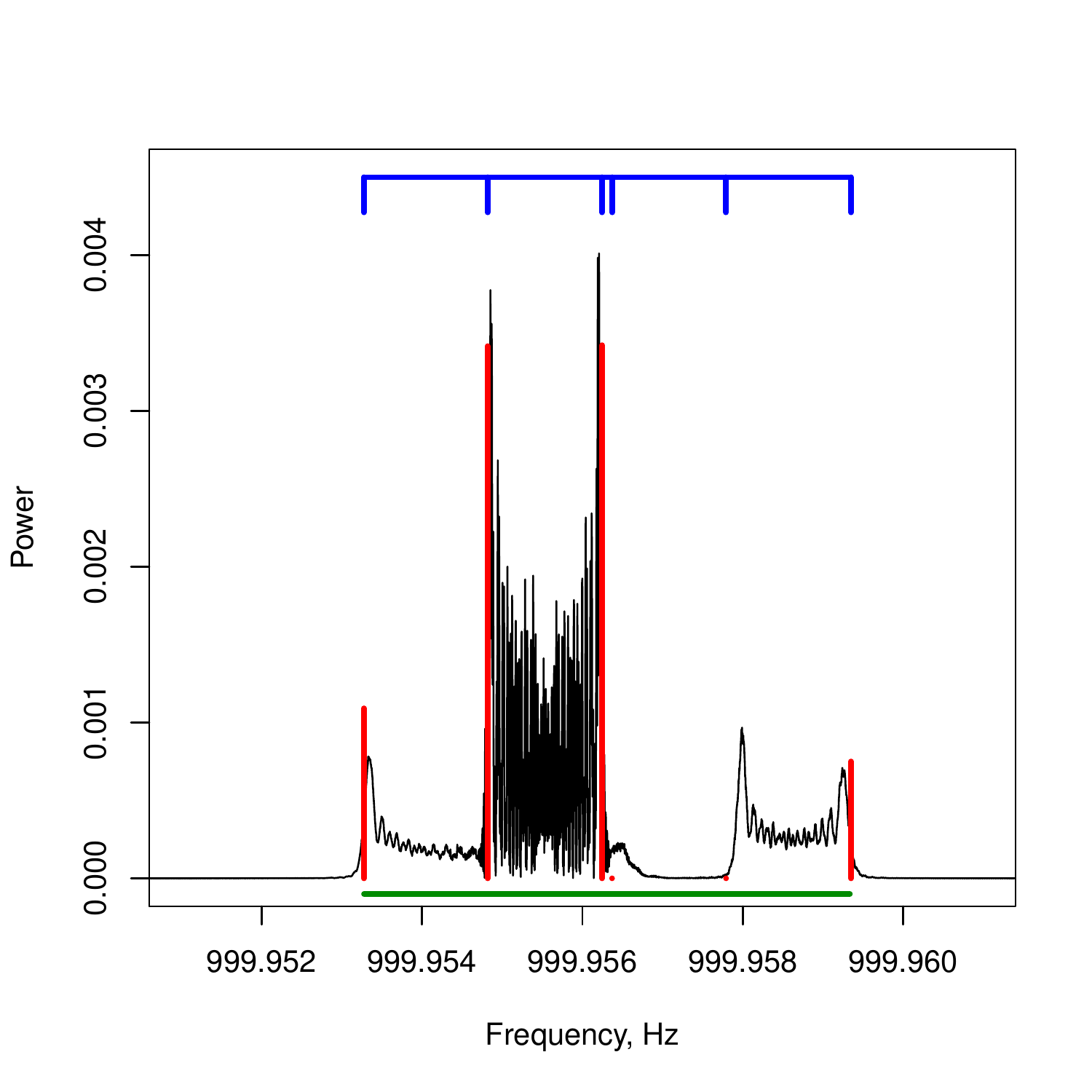}
 \caption{Example 3-day Fourier transform of 1000\,Hz monochromatic signal from source at right ascension $0^\circ$ and declination $0^\circ$. The green line at the bottom of the plot shows spectrum support region estimated using equation \ref{eqn:spectrum_support}. The blue line at the top shows peak locations estimated using equation \ref{eqn:peak_locations}. The red lines show peak strength estimated using equation \ref{eqn:peak_height2}. The Fourier transform was computed assuming the data stretch had a 30 hour gap in data for LIGO Hanford interferometer. The 3-day segment started at GPS 1160657033. The gap started 10 hours later.}
\label{fig:spectrum_support1_gap}
\end{center}
\end{figure}

This simple formula works surprisingly well. An illustration is given in Figure \ref{fig:spectrum_support1_gap}. Here we marked both peak locations and their strength. Also for this example we introduced a gap of 30 hours, demonstrating the ability to handle non-contiguous data.

\section{Spectrum shape algorithm}

\begin{figure}[htbp]
\begin{center}
\fbox{
\begin{minipage}{0.95\linewidth}
\begin{enumerate}
\item The input to the algorithm is a series of timestamps $\{s_i\}_{i=1}^{\tilde N}$ in the local interferometer frame, 
    as well as the computed times $\{t_i\}_{i=1}^{\tilde N}$ in solar system barycenter frame.
\item Compute the fit $t_i=\sum_{n=1}^N g_n s_i^n + A \cos(\omega_1 s_i+\phi_1)$.
      $N$ can be taken as $3$ for time intervals shorter than 3\,days. $\omega_1$ is the rotational frequency correspondingly to Earth sidereal period.
\item Compute times of zero frequency derivative  and their corresponding stationary frequencies (see equations \ref{eqn:stationary_points} and \ref{eqn:peak_locations}). These give peak locations.
\item Compute times of zero second frequency derivative (inflection points, equations \ref{eqn:inflection_locations1} and \ref{eqn:inflection_locations2}).
\item Estimate peak amplitudes with formula \ref{eqn:peak_height2}, where $t_0$ and $t_1$ are nearest inflection points or data boundaries.
\item{The output of the algorithm consists of estimated peak frequencies and amplitudes}
\end{enumerate}
\end{minipage}
}

 \caption{Algorithm used to compute the shape of Fourier transform}
\label{fig:algorithm}
\end{center}
\end{figure}

The analysis detailed in the previous section can be condensed into the algorithm for determining Fourier transform spectral shape (Table \ref{fig:algorithm}).

At the start of the algorithm we compute a sequence of times relative to Solar System barycenter. This could be done exactly, or as an approximation. For example, one can compute these times for a relatively coarse grid on the sky and then use a suitable method, such as \cite{Sauter2019} to interpolate between locations.

Once this time series has been obtained it can be fitted to the formula
\begin{equation}
t_i=\sum_{n=1}^N g_n s_i^n + A \cos(\omega_1 s_i+\phi_1)
\end{equation}
over an interval matching the coherence length of the Fourier transform. Long stretches of data are best analyzed using overlapped intervals. A straightforward speedup is to interpolate the fits from those computed on a coarse grid. 

Also, iteration over signal waveforms with the same sky location but with different frequency drift is achieved by direct modifications of coefficients $g_k$.

With the fit in hand, it is straightforward to find locations of stationary points $t_a$ and inflection points $t_b$ (equations \ref{eqn:stationary_points}, \ref{eqn:inflection_locations1}, \ref{eqn:inflection_locations2}). 

Now the frequencies of the peaks are given by formula \ref{eqn:peak_locations} and peak height is computed using formula \ref{eqn:peak_height2}. 

The computed spectrum shape can be used to understand the dwell time of signal waveform and used to characterize and mitigate the influence of detector artifacts - either after analysis by removing outliers coincident with detector lines, or during the analysis by decreasing weight of segments with larger peak heights.

\section{Performance}

The formulas \ref{eqn:peak_locations}, \ref{eqn:peak_height1}, \ref{eqn:peak_height2} are very efficient compared to computing Fourier transform from scratch or to numerically integrating equation \ref{eqn:time_lattice}. 

For the direct problem of determining peak locations and amplitudes, any method would need to compute timestamps first. The computing cost of the Fourier transform is similar or larger to the cost of computing timestamps - such algorithms have theoretically steeper scaling of $C {\tilde N} \log(\tilde N)$ of computational effort compared to number of $\tilde N$ of input data points. In practice the running time of the Fourier transform is strongly influenced by implementation efficiency or, in other words, constant $C$. A general purpose library algorithm will not be as efficient as hand-tuned implementation for fixed input size.

Our algorithm replaces the Fourier transform with a fit of computed timestamps, that is easy to optimize taking advantage of vector arithmetic. Moreover, for large parameter searches we would need to perform this computation repeatedly for a range of signal parameters, in particular frequency and frequency derivative. In such a situation, computing Fourier transform over and over again is very expensive. Our algorithm computes peak timings with a simple formula, and can be used to translate signal parameter range into the range of peak locations and amplitude.

The inverse problem of determining signal parameters that correspond to a known instrumental line is even harder to solve with brute force Fourier transform, as it will require to sample a large grid to check for coincidence of computed peaks with the line. As expected an analytic formula is much faster.

The efficiency of the algorithm is contingent on the validity of the underlying model. To test how well this model fits the data we made a study using numerically computed timings. 

A coarse sky grid of 182 points was used for this study. The points on the grid were arranged in $18^\circ$ increments in declination and right ascension. Only one value of right ascension was used for equatorial poles with declination of $\pm 90^\circ$.

For each point in the sky grid we generated $17520$ Solar System Barycenter timings using routines from LAL library \cite{LALlibrary}. The timing started at GPS $1160657033$ with $0.5$\,hour increments. Separate datasets were generated for LIGO Hanford and Livingston interferometers.

Using this dataset we tested fit to the single harmonic model:
\begin{equation}
t'=\sum_{n=0}^{N} g_n \frac{t^n}{n!}+A \sin(\omega_{\textrm{s}} t+\phi_1)
\end{equation}
where $t$ is the time in the detector frame of reference, $t'$ is the time at Solar system barycenter and $\omega_{\textrm{s}}$ is the angular frequency corresponding to Earth sidereal rotation period. All other coefficients were fitted.

For each point in the sky the entire set of timestamps was separated in 3-day stretches, with nearby stretches overlapped by 1.5\,days. Each stretch was fitted using at most cubic terms $N=3$. The absolute worst residual maximized over all stretches and all sky points was $13.2$\,$\mu$s .

A similar procedure was performed for 6-day stretches, this time increasing the number of polynomial terms to $N=4$ and using a 3-day overlap. The absolute worst residual was $24.7$\,$\mu$s. 

As the typical signals studied in continuous gravitational wave searches go up to $2$\,kHz the precision of the fit is sufficient to apply results described in this paper. 

\section{Application to analysis of outlier at 1891.76 Hz}

\begin{figure}[htbp]
\includegraphics[width=3.3in]{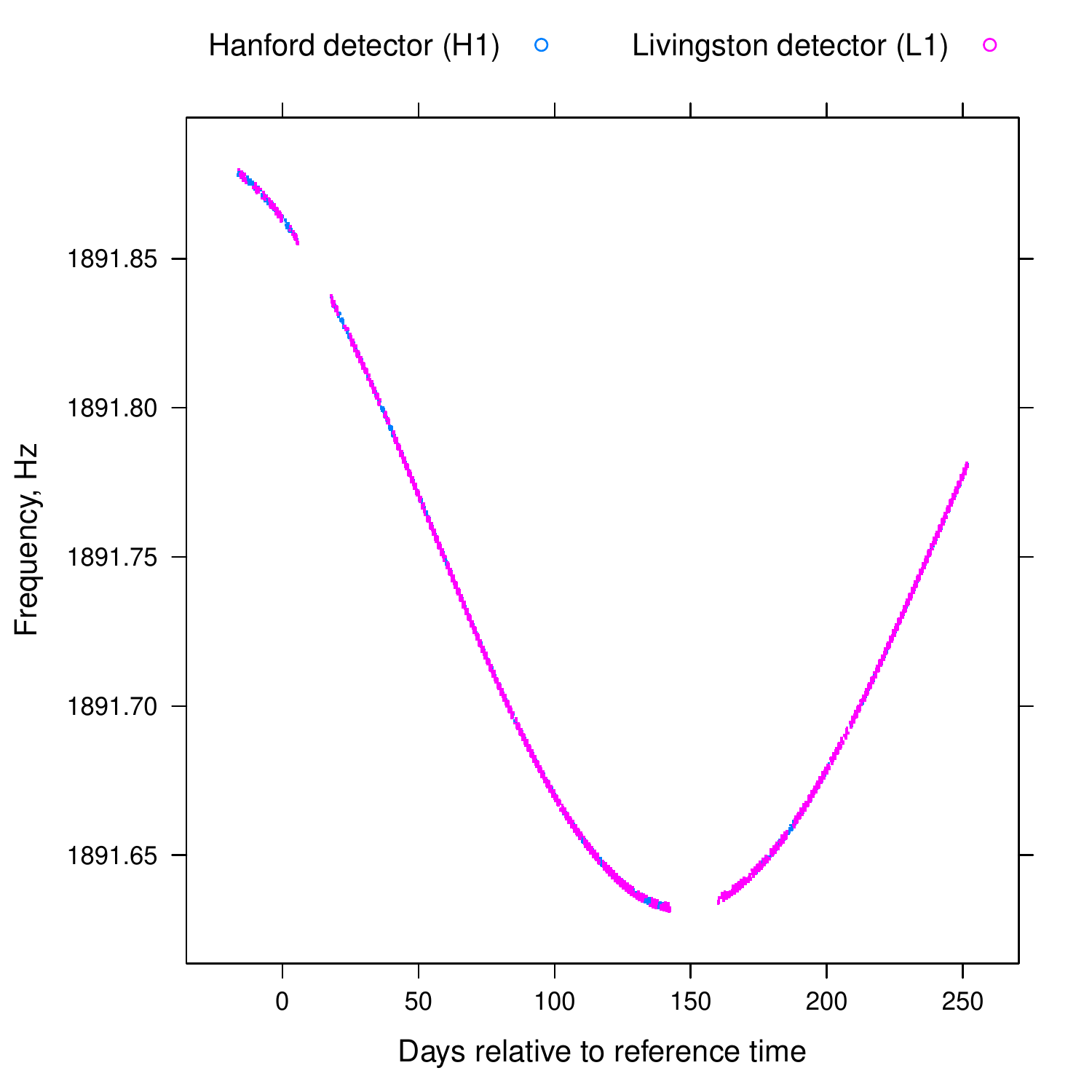}
\caption[Outlier map]{
\label{fig:outlier_evolution}
Apparent frequency of a signal with parameters equal to those of the outlier at 1891.76\,Hz, at the detectors.  The difference in Doppler shifts between interferometers is small compared to the Doppler shifts from the Earth's orbital motion. The reference time is at GPS epoch $1183375935$.
}
\end{figure}
\begin{figure}[htbp]
\includegraphics[width=3.3in]{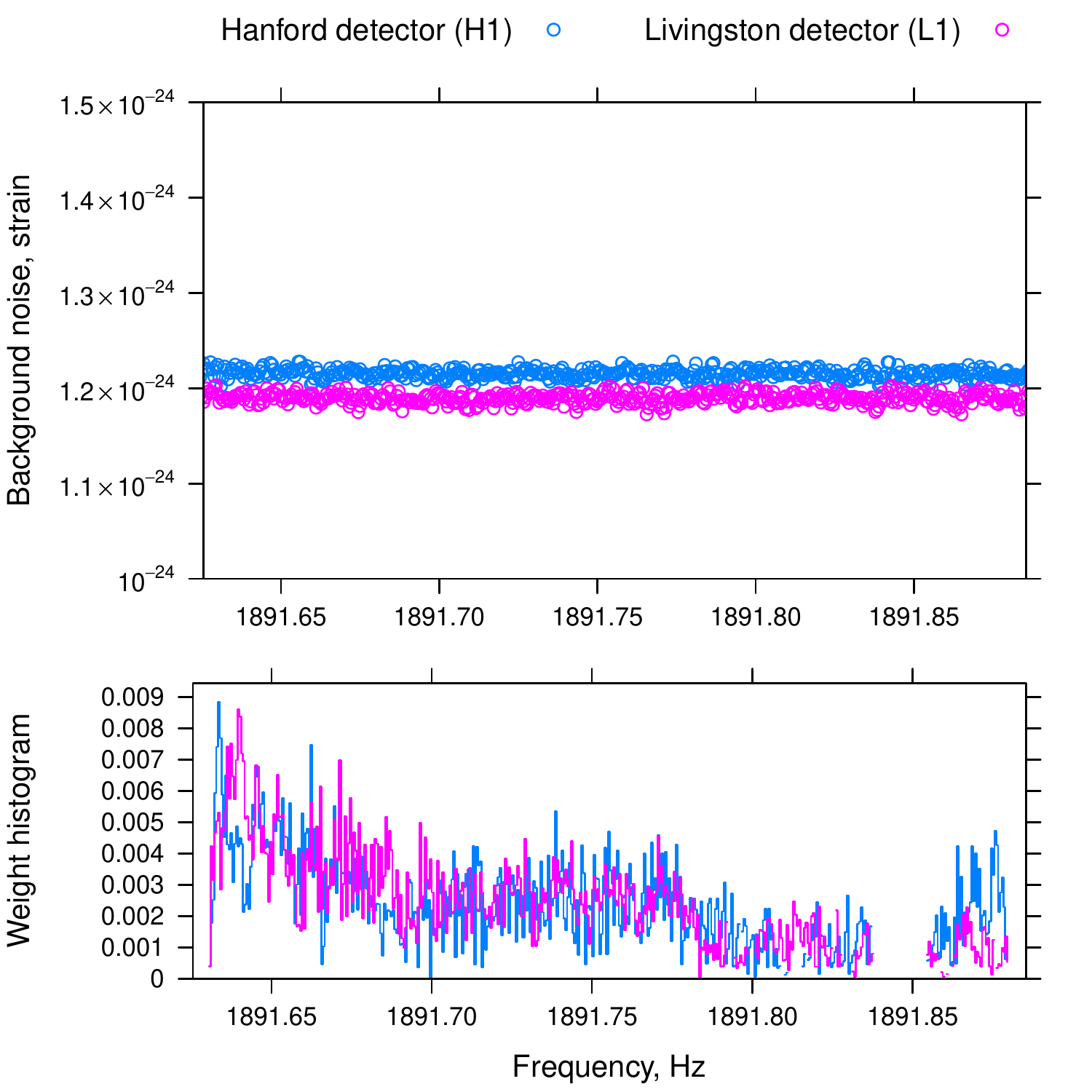}
\caption[Outlier spectrum]{
\label{fig:outlier_hist}
The top plot shows the average amplitude spectral density around the frequency of the outlier at 1891.76\,Hz. The bottom plot shows how much the power in each frequency bin would contribute, over the course of the O2 run, to the total power estimated by the Falcon pipeline for a signal with the parameters of the outlier at $\approx$ 1891.8 Hz. High-weight values correspond to bins with greater contribution and the sum of all the weights is 1 for each curve. The gap corresponds to a break in the O2 run and matches the gap in the frequency evolution plot (Figure \ref{fig:outlier_evolution}).
}
\end{figure}

The outlier at 1891.75674\,Hz identified in recent search \cite{O2_falcon2} is an excellent target to illustrate application of our method. If it is of astrophysical origin it would likely correspond to the fastest rotating neutron star found to date. This outlier had a frequency drift of $-8.22\times 10^{-12}$\,Hz/s and rather wide frequency evolution as illustrated in Figures \ref{fig:outlier_evolution} and \ref{fig:outlier_hist} (reproduced from \cite{O2_falcon2}). 

As shown on Figure \ref{fig:frequency_evolution} the outlier spanned almost 0.25\,Hz because of large Doppler shifts due to orbital motion. Because the O2 science run spanned 9.5\,months there is a region with stationary frequency. For this outlier it happens to overlap a gap in data taking. 

Figure \ref{fig:outlier_hist} shows interferometer spectrum in the top panel. It was generated without applying any outlier specific corrections such as Doppler shifts or amplitude modulation, but using the same noise-weighting scheme. The individual points correspond to individual bins in 1800\,s long Hann-windowed Fourier transforms of the underlying data. This is a rather coarse resolution compared to the 6-day long coherence length used in the last stage of outlier followup in \cite{O2_falcon2}. However, the plot does show an absence of large detector artifacts - such as narrow lines that occur at other frequencies.

In order to better understand this outlier it is desirable to figure out whether some coherent detector noise is being masked by shot noise background. To do this we examine the locations of stationary peaks of our outlier. As it has arisen from a search over wide parameter space, the presence of instrumental noise source should force those peaks to cluster so as to raise the signal-to-noise ratio of the outlier. 

\begin{figure*}[htbp]
\includegraphics[width=7.2in]{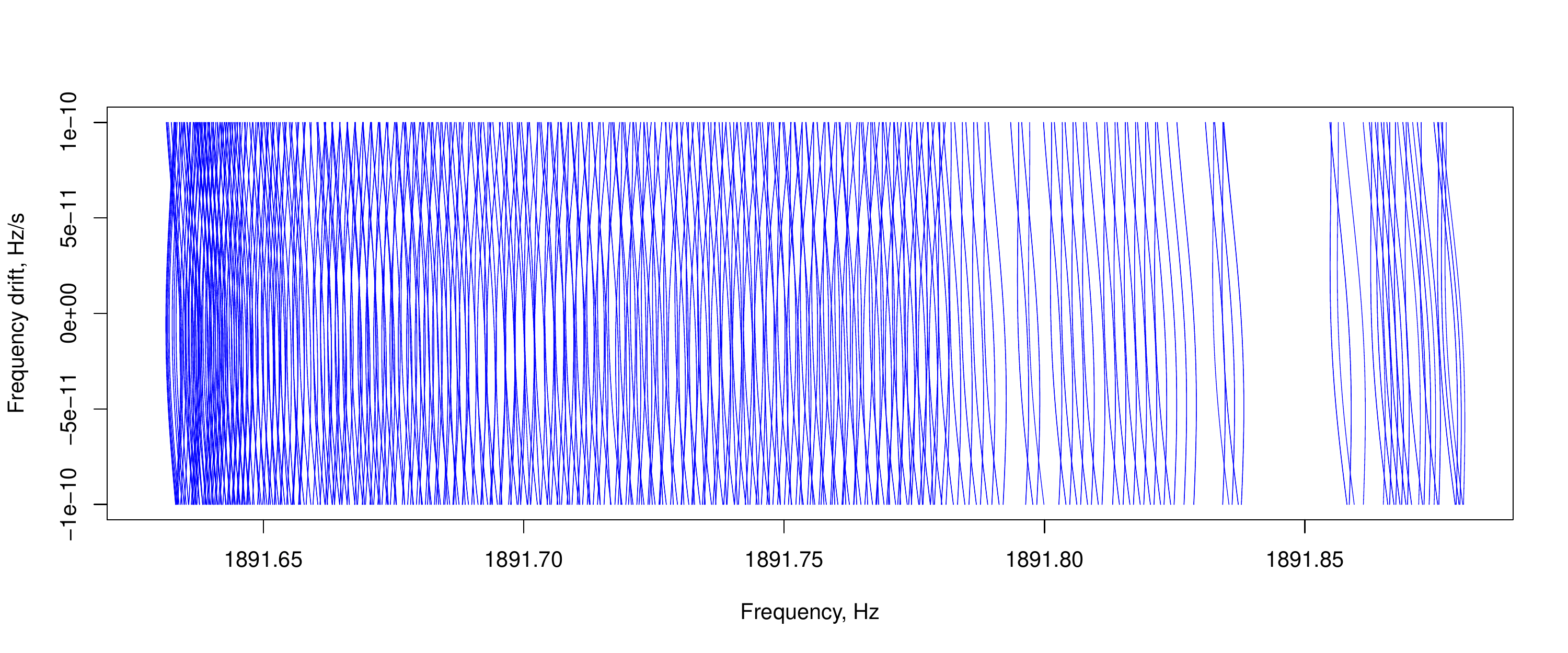}
\caption[Outlier peaks]{
\label{fig:outlier_peaks}
Dependence of stationary peak frequency on frequency drift for outlier at 1891.76\,Hz using data corresponding to LIGO Livingston Observatory (LLO). The dense area on the left of the plot corresponds to the frequency minimum seen in Figure \ref{fig:outlier_evolution}
}
\end{figure*}

\begin{figure*}[htbp]
\includegraphics[width=7.2in]{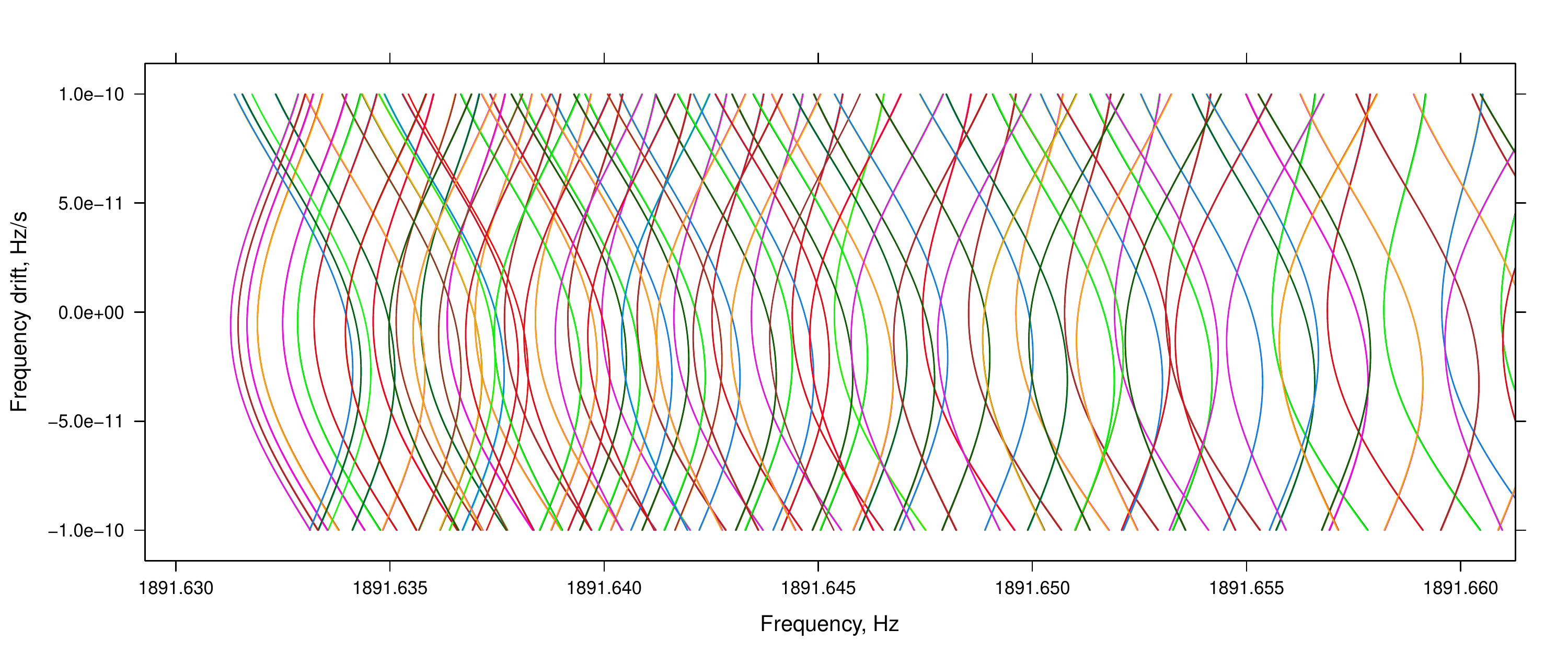}
\caption[Outlier peaks zoomed]{
\label{fig:outlier_peaks_zoomed}
Dependence of stationary peak frequency on frequency drift for outlier at 1891.76\,Hz. We show a magnified view of lower band of outlier frequency evolution. Different colors identifiy individual peaks. }
\end{figure*}

\begin{figure*}[htbp]
\includegraphics[width=7.2in]{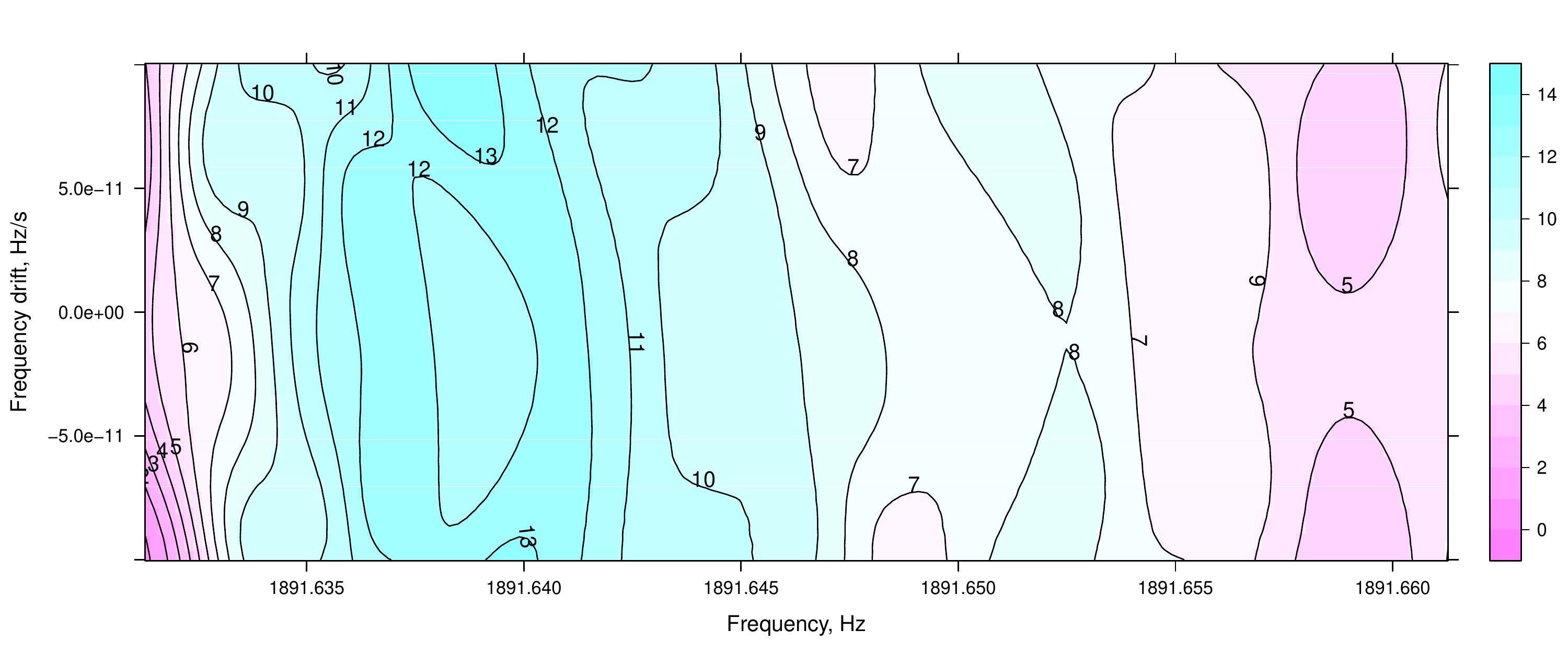}
\caption[Outlier peaks density map zoomed]{
\label{fig:outlier_peaks_density_zoomed}
Density of stationary peaks for outlier at 1891.76\,Hz. We show a magnified view of low-frequency band. The density was computed using the Gaussian kernel with bandwidth of 1\,mHz.}
\end{figure*}

Figure \ref{fig:outlier_peaks} shows how peak frequencies change with outlier frequency drift for LLO detector at LIGO Livingston observatory. Having analytic formulas was essential to producing this plot, as otherwise we would need to iterate over individual frequency drift values - a procedure that is difficult to do even on a large cluster because of large demands on storage bandwidth of existing software. 

We observe a fairly even distribution of peak on Figure \ref{fig:outlier_peaks}, except for gaps due to gaps in O2 science run and a dense area on the left corresponding to outlier frequency minimum. 

A magnified view of peak near outlier frequency minimum is shown on Figure \ref{fig:outlier_peaks_zoomed}. We used different colors to separate individual peak lines. We observe that as frequency drift varies most peaks move in the same direction, with small variations between individual peaks. There is no single area where many peaks cross that would allow a sharp instrumental line to have undue influence.

However, it could be that there is a wider hidden instrumental line and the outlier takes advantage of that by increasing peak density where the line is. To study this we made plots of peak density, one of which - a magnified view of low frequency band - is shown in Figure \ref{fig:outlier_peaks_density_zoomed}. We observe that there is nothing special about the frequency drift corresponding to the outlier at 1891.76\,Hz, and, in fact, there are nearby areas of larger frequency drift and larger peak density. 

Similar plots for LHO detector at LIGO Hanford observatory are similar, except for showing smaller peak density. Our conlusion from these studies is that it is highly unlikely that the outlier at 1891.76\,Hz was caused by a stationary detector artifact.

\section{Conclusions}

The question of identification of continuous wave outliers to detector disturbances is of utmost importance in separating astrophysical signals from detector artifacts. 

In this paper we analyze the shape of the Fourier transform of continuous wave gravitational wave signal and present simple formulas to compute peak heights and locations arising from features in frequency evolution of gravitational wave signal.

While our focus was on understanding Fourier transform of a gravitational wave signal, the formulas and the analysis presented here can be applied to any signals of this form.

We apply our formulas to investigate the outlier at 1891.76\,Hz \cite{O2_falcon2} and find that this outlier is unlikely to be induced by a stationary detector artifact.

\section{Acknowledgments}\label{sec:Acknowledgments}
S. R. Valluri would like to acknowledge The Natural Sciences and Engineering Research Council of Canada (NSERC) for a Discovery Grant during the course of this work. We would also like to thank Sheel Patel for a thorough proof read of the paper.

\end{document}